\documentclass[a4paper]{jpconf}
\usepackage{graphicx}
\begin{document}
\title{A Security Monitoring Framework For Virtualization Based HEP Infrastructures}

\author{A. Gomez Ramirez\textsuperscript{1}, M. Martinez
Pedreira\textsuperscript{2}, C. Grigoras\textsuperscript{2}, L.
Betev\textsuperscript{2}, C. Lara\textsuperscript{1} and U. Kebschull\textsuperscript{1} for the ALICE Collaboration}

\address{\textsuperscript{1}Infrastructure and Computer Systems for Data
Processing (IRI), Goethe-University Frankfurt}
\address{\textsuperscript{2}CERN, Geneva, Switzerland}

\ead{andres.gomez@cern.ch}

\begin{abstract}
High Energy Physics (HEP) distributed computing infrastructures require
automatic tools to monitor, analyze and react to potential security incidents. 
These tools should collect and inspect data such as resource consumption, 
logs and sequence of system calls for detecting anomalies that indicate the presence of a malicious
agent. They should also be able to perform automated reactions to attacks
without administrator intervention. We describe a novel framework that accomplishes these
requirements, with a proof of concept implementation for the ALICE experiment at
CERN. We show how we achieve a fully virtualized environment that improves 
the security by isolating services and Jobs without a significant performance
impact. We also describe a collected dataset for Machine Learning based Intrusion
Prevention and Detection Systems on Grid computing. This dataset is composed of
resource consumption measurements (such as CPU, RAM and network traffic), logfiles 
from operating system services, and system call data collected from production 
Jobs running in an ALICE Grid test site and a big set of malware. This malware 
was collected from security research sites. Based on this dataset, we will 
proceed to develop Machine Learning algorithms able to detect malicious Jobs.
\end{abstract}

\section{Introduction}
Frequently in HEP computing, and also in general purpose Grid computing, user
supplied code and data are deployed and executed in farms around the world, 
while the exact location is normally irrelevant. This allows
scientists from many areas beyond physics to use huge computational
power to solve complicated scientific problems, such as weather
modeling, brain simulation, among others. However it also creates
cyber-security challenges. Operators and administrators need
tools to monitor for security incidents. User code and data should be
isolated from different users, also from the physical computers and
networks, in order to restrict access to sensitive elements in the organizations.
\par
We propose a novel paradigm and developed a framework that focus on
protecting and monitoring user payload execution. Moreover, it enforces
isolation in the environment in such a way that Jobs cannot access sensitive 
resources. This tool enables the Job behavior analysis in
order to detect possible intrusions. This is accomplished by collecting and 
processing data generated by Jobs such as logs, system calls and resource
consumption data. Traditional Intrusion Detection and Prevention Systems
(IDPS) perform attack detection by using fixed rules based on signatures,
identical to traditional monitoring systems. Therefore, we employ Machine Learning (ML) 
to overcome the mentioned drawbacks, achieving generalization among attack
variants. Currently there is no tool that provides isolation, while monitoring 
security incidents by ML algorithms in Grid computing \cite{Gomez}.
\par
The authors have defined a threat model that guides the design and
implementation of the described framework \cite{Gomez}. It is devised to detect
attackers in the protected system trying actions like the following:

\begin{itemize}
  \item Exploit unknown or unfixed software/hardware vulnerabilities.
  \item Listen to user network traffic to gather sensitive clear text
  information.
  \item Perform a 'man in the middle' attack.
  \item Tamper other user Jobs.
  \item Escalate privileges.
  \item Access sensitive server configuration data.
\end{itemize}

\par
\begin{figure}[h]
\begin{center}
\begin{minipage}{100mm}
\includegraphics[width=100mm,scale=1.0]{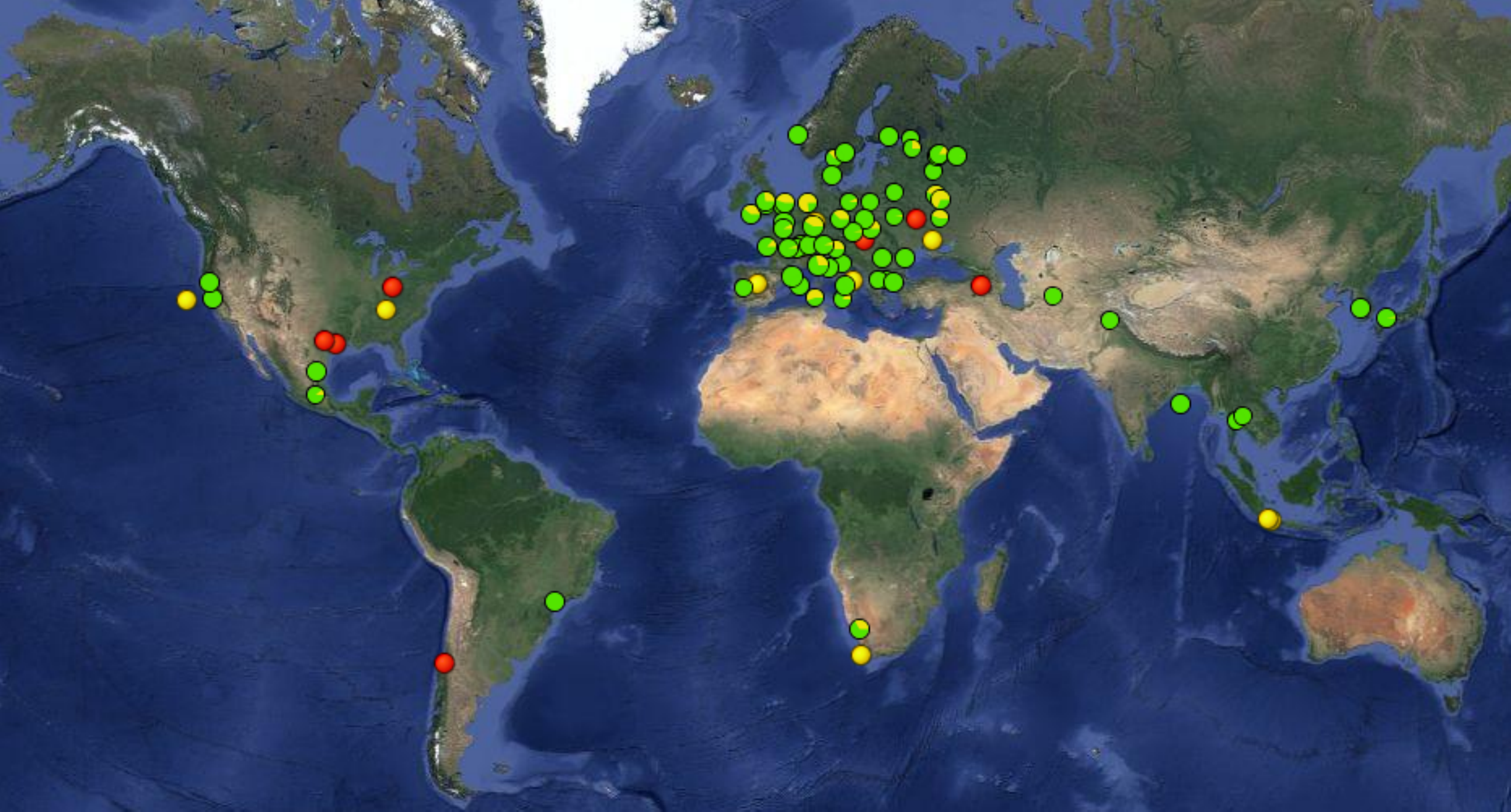}
\caption{\label{fig:figure1}ALICE Grid computing farms aroud the world.}
\end{minipage}\hspace{2pc}%
\end{center}
\end{figure}

As a proof of concept we are implementing the described framework for the ALICE
Grid at CERN. ALICE (A Large Ion Collider Experiment) is a dedicated Pb--Pb
detector designed to exploit the physics potential of nucleus-nucleus interactions at the Large 
Hadron Collider at CERN \cite{AliceProposal,AliceExperiment}. The ALICE
experiment has developed the ALICE production environment (AliEn) \cite{alien},
which implements many components of the Grid technologies that are
needed to analyze HEP data. Through AliEn, the computing centers that
participate in the ALICE Grid can be seen and used as a single entity. Any
available node executes Jobs and file access is transparent to the user, 
wherever in the world a file might be \cite{GridTDR}.
Figure~\ref{fig:figure1} shows a picture of the ALICE Grid.

\par

This document is organized as follows: section~\ref{isolation} introduces
a security by isolation strategy for a distributed system.
Section~\ref{MonitoringMining} explains a method for collecting relevant data
from the isolated infrastructure. Section~\ref{MachineLearning} shows how the
collected data can be used to determine the security status of the system. 
In section~\ref{IntrusionPreventionDetection} we detail our Intrusion 
Detection and Prevention model. Section~\ref{ProofConcept} summarizes the
current state of the project and the challenges faced in the design and implementation of the 
desired methodology. Finally, section~\ref{conclusions} gives conclusions on
the work done.

\section{Security by isolation}
\label{isolation}
Security by isolation enforces application space separation \cite{isolation}.
The idea is that, if one process is compromised and utilized to attack the
entire system, other components can stay untouched.
Several technologies provide secure isolation. Virtual Machines (VM) 
and Linux Containers (LC) are very popular examples.

\subsection{Linux containers}
LC are an extension of the virtual memory concept to allow the isolation of
network interfaces, the PID tree and mount points \cite{containers}.
Separation of containers from the rest of the system is enforced by the Kernel, 
so they can not affect the host or other containers. LC technology uses 
namespaces and Cgroups \cite{namespaces} to have a private view of the system
and a limited resource assignment.

\par
Containers provide a set of advantages over VM. They are
lightweight and fast, boot in milliseconds and have just a few MB of intrinsic
disk and memory usage. It has been shown \cite{IBM}, that they provide a better
performance than VMs. Commonly, VM are used in Grid and Cloud computing to
achieve isolation, however LC performance and comparable security features make
it a suitable alternative\cite{containersincloud, adversary}.

\subsection{Proposed isolation architecture}
\begin{figure}[h]
\begin{center}
\begin{minipage}{100mm}
\includegraphics[width=100mm,scale=1.5]{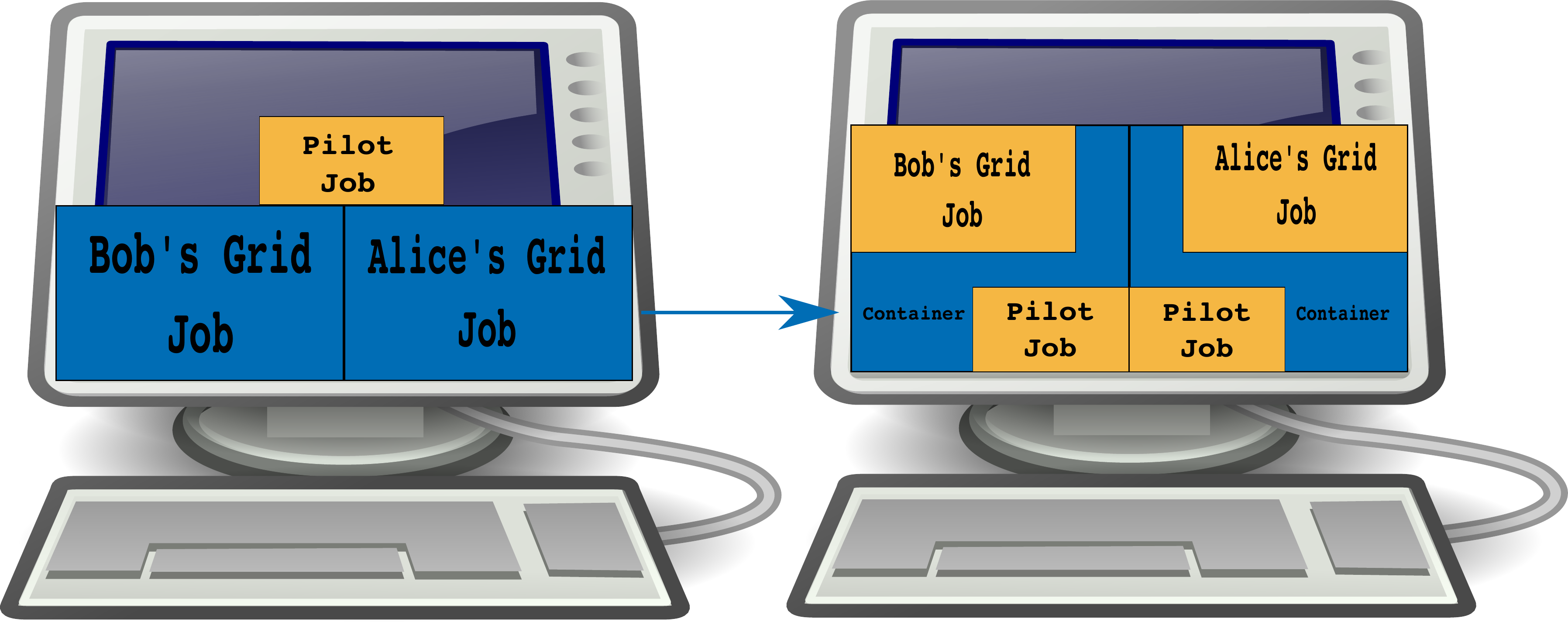}
\caption{\label{fig:figure2}Desired isolation scenario.}
\end{minipage}\hspace{2pc}%
\end{center}
\end{figure}

We propose the usage of LC to enforce HEP Grid site user isolation, also
extensible to broader scientific computing and clouds. To achieve this we require
a batch Job orchestrator allowing the execution of user processes in containers on computing
clusters. Section~\ref{ProofConcept} gives further details about this
requirement and the selected solution. As shown in Figure~\ref{fig:figure2},
we switch from an environment without isolation, where Jobs have access to the
server and other user Jobs, to an environment where Jobs run in their one
process space, without access to other Jobs or sensitive components.

An isolated environment for Job execution is not enough. Jobs could
still perform several kind of attacks or not allowed activities such as,
Distributed Denegation of Service (DDoS), Bitcoin mining, malware/botnet
hosting, among others. Consequently, we need to monitor the activity and detect
incidents inside the LC, as described in the next sections.

\section{Monitoring data mining}
\label{MonitoringMining}
HEP distributed computing systems use continuous automated monitoring 
to help administrators to find and fix situations affecting the normal
operation \cite{monitoring}. The resulting monitoring data can be used to find
or even predict software and hardware failures. It is a valuable source of security
information as well. In this document we focus on the measurement of metrics
related to the batch Jobs being submitted to the distributed system. There are
several relevant metrics that we can collect, for instance:

\begin{itemize}
  \item Job and system logs.
  \item System call sequence.
  \item Resource usage data (such as CPU, RAM and network traffic).
\end{itemize}

Furthermore our goal is to chose the best information about Job behavior without
affecting the habitual performance. We decided to employ data mining and
intelligent algorithms, given their ability to find correlations and analyze 
trends in big datasets \cite{datamining}, in order to provide a better
understanding  of security related events.

\section{Machine learning based security monitoring}
\label{MachineLearning}
Machine Learning is a set of mathematical models that simulate the
human learning abilities\cite{svm1, svm2}. In the context of Intrusion 
Detection, ML helps analyzing big amounts of data by learning the 
expected behavior and identifying abnormal situations. Traditional industrial
IDS use rather fixed rules and search for known attack signatures. However
they have problems when unknown or slightly different intrusion 
methods are used \cite{datamining}. We have selected supervised training to
analyze the collected data. In supervised training, a set of already classified
and tagged data (training dataset) is used to model a function (for example a
Neural Network) in order to make it able to classify new unseen data
(test dataset) \cite{bishop}.

\begin{figure}[h]
\begin{center}
\begin{minipage}{100mm}
\includegraphics[width=100mm,scale=1.5]{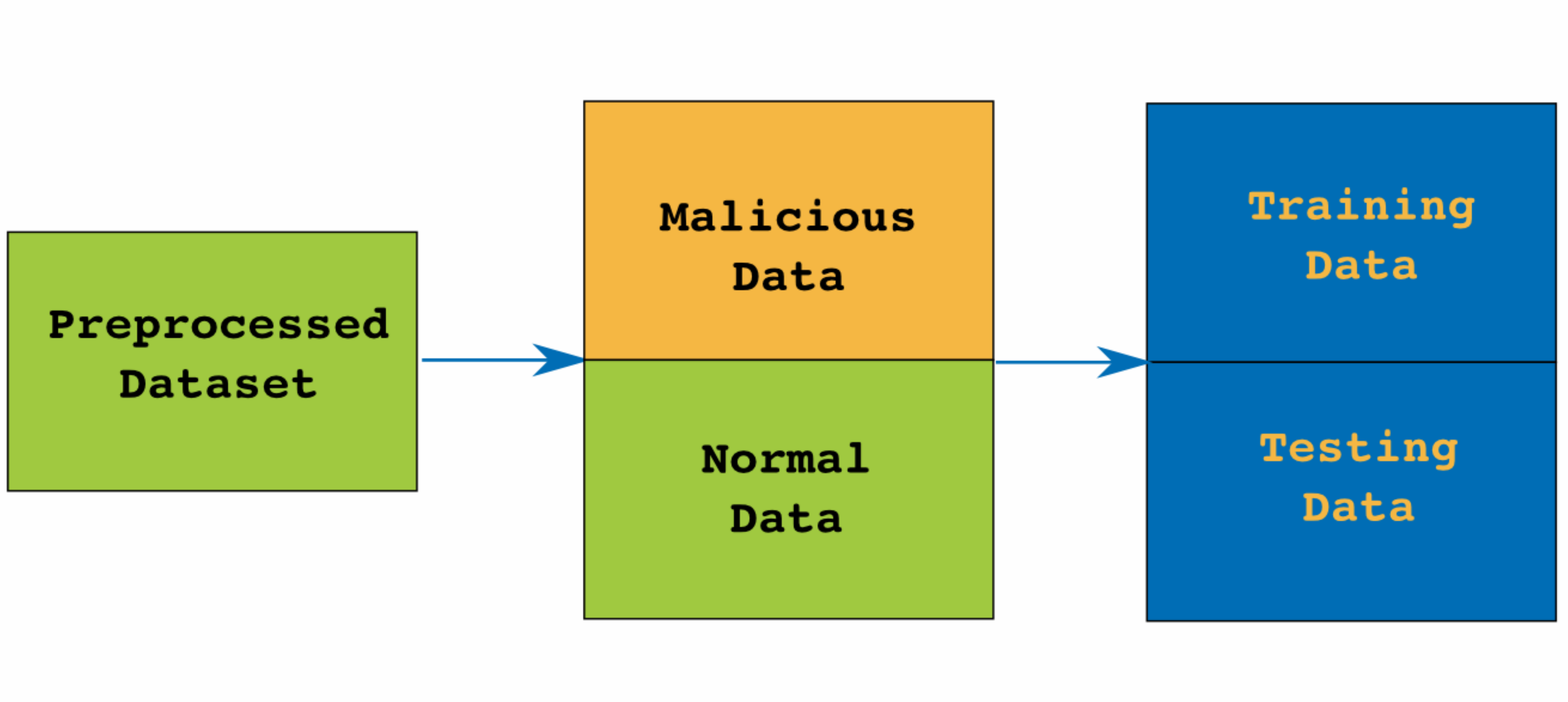}
\caption{\label{fig:figuremon}Monitored data gathering and processing.}
\end{minipage}\hspace{2pc}%
\end{center}
\end{figure}

\subsection{Training dataset}
\label{training}
We have collected a training and testing dataset by gathering monitoring data
from production ALICE Grid Jobs (Figure~\ref{fig:figuremon}). Additionally we
have executed a big set of Linux malware samples. The data on the first part 
is tagged as normal data and the second as malicious. This dataset is utilized 
to compare several Machine Learning algorithms to find the one that gives the best accuracy. 
Following is the list of ML algorithms selected. They will be tested to define
which one gives the best accurate results for our dataset:

\begin{itemize}
  \item Support Vector Machines.
  \item Multilayer Neural Networks.
  \item Recurrent Neural Networks.
\end{itemize}

\par
\begin{figure}[h]
\begin{center}
\begin{minipage}{150mm}
\includegraphics[width=150mm,scale=1.5]{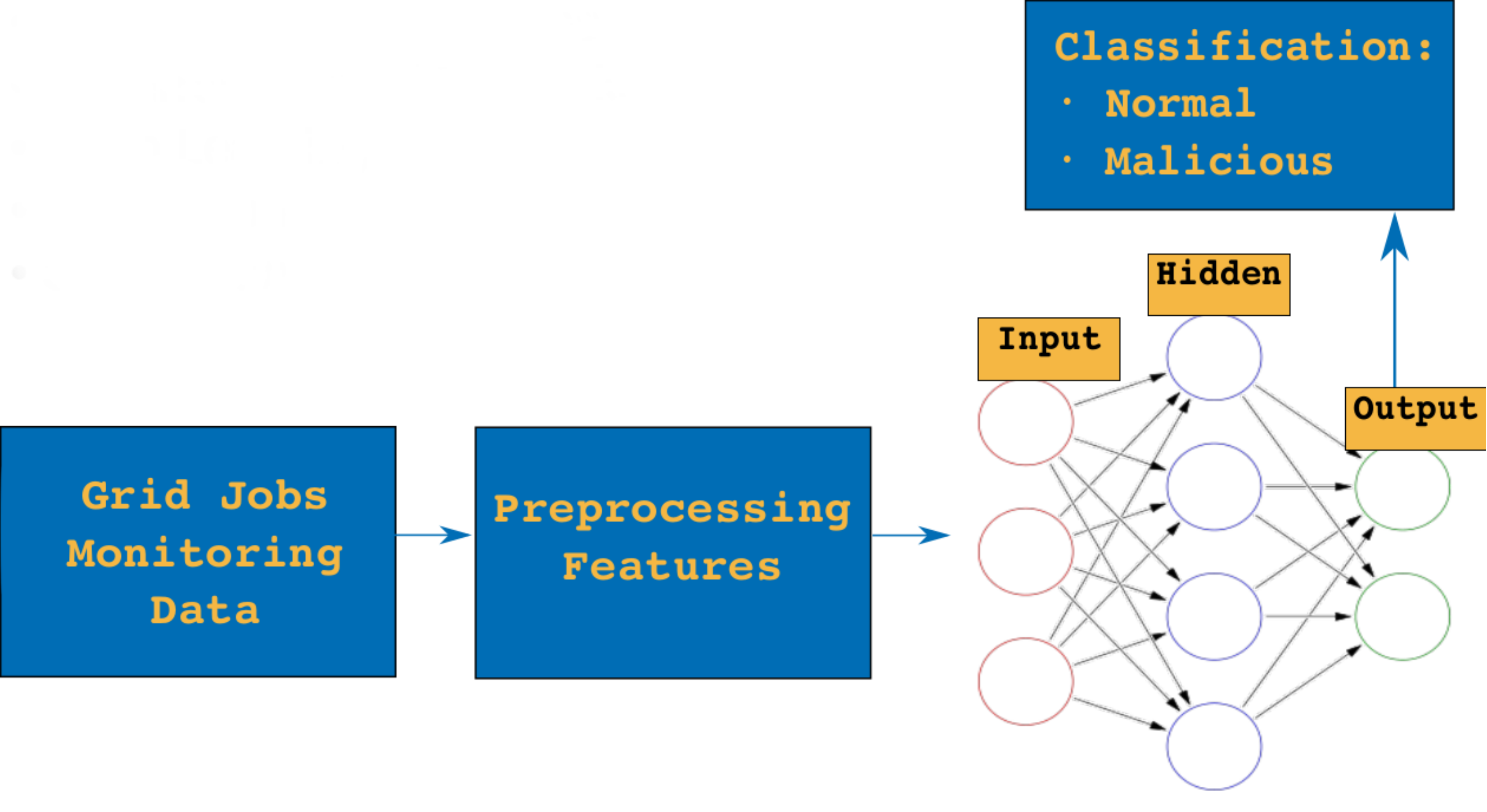}
\caption{\label{fig:figureflow}Proposed architecture.}
\end{minipage}\hspace{2pc}%
\end{center}
\end{figure}

Figure~\ref{fig:figureflow} shows a scheme of the proposed architecture for the 
ML usage.

\section{Intrusion prevention and detection}
\label{IntrusionPreventionDetection}
\begin{figure}[h]
\begin{center}
\begin{minipage}{150mm}
\includegraphics[width=150mm,scale=1.5]{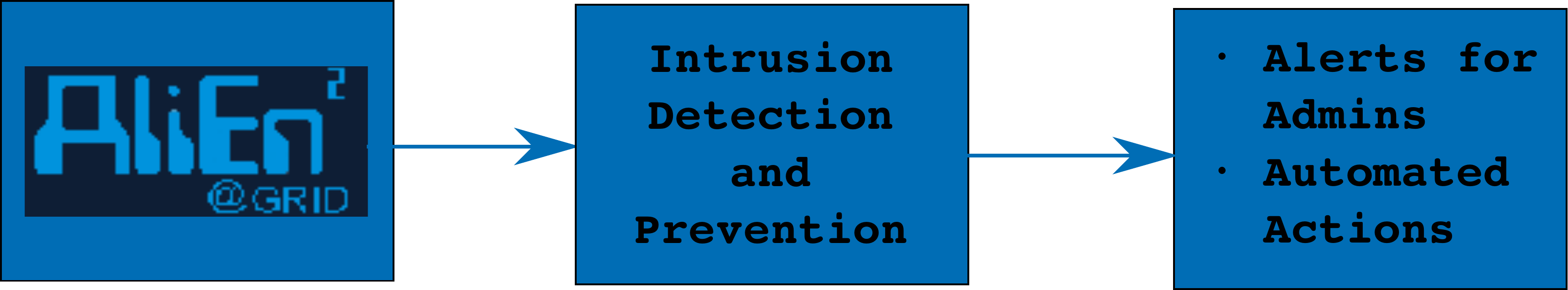}
\caption{\label{fig:figureids}Proposed Intrusion Detection in the worker nodes.}
\end{minipage}\hspace{2pc}%
\end{center}
\end{figure}

When analyzing Job monitoring data, our goal is finding security 
incidents. This security incidents can be found by analyzing anomalies in the
system, things that go beyond the common state, probably caused by malicious
software. Besides, even if our execution environment is sandboxed, there are many possible 
attacks that can still affect the distributed infrastructure. If a user's
Job is misbehaving, the proposed framework should raise an alarm and perform 
predefined actions, for instance terminate the malicious processes. Figure~\ref{fig:figureids} 
shows the desired implementation of the proposed system regarding Intrusion Detection.

\subsection{Challenges}
\label{Challenges}
Improving the provided security should not impact the system
performance. This is especially important in HEP computing. On the other
hand, we also need innovative ways to analyze the trace and log data in 
an efficient way. Another important challenge is to reduce the amount of 
false positives and false negatives, since the system administrators rely 
on the accuracy of the security monitoring framework.

\section{Proof of concept and testing environtment}
\label{ProofConcept}
So far we have already deployed a testing ALICE Grid site based on AliEn 
\cite{alien} in a local Linux cluster, with five Ubuntu 14.04 nodes. In order to 
orchestrate and run the Jobs inside Linux Containers we have tested three 
different tools that offer such functionality:
 
\begin{itemize}
  \item Kubernetes \cite{kubernetes}.
  \item Apache Mesos \cite{mesos}.
  \item Docker Swarm \cite{swarm}.
\end{itemize}

\par
At the end we have decided to work with Docker Swarm, because it allows to
carry out the simplest deployment, which is an important requirement for our
research environment. We use Docker \cite{docker} as LC engine, with Centos 6
\cite{centos6} container images. We have developed AliEn
interfaces for the mentioned batch systems. CVMFS \cite{cvmfs} is installed on
the hosts and shared as a volume inside the AliEn container to allow access to 
HEP libraries. Currently we execute one Job per container. This is useful to 
increase the traceability between different Jobs. Also, this is the natural
micro service model for LC.

\par
As a monitoring infrastructure for collecting data from normal Grid Jobs we
have Prometheus \cite{prometheus} and Sysdig \cite{sysdig}. Prometheus allows
to take resource usage data directly from containers and collect it via a
RESTful interface. Sysdig enables to capture system calls in Linux OS in a fast
a reliable way. We have develop a custom Python library to integrate these tools and make them fit our 
needs. This infrastructure has been utilized for the execution and measurement
of ALICE production Jobs, that are tagged as normal Jobs.

\par
\begin{figure}[h]
\begin{center}
\begin{minipage}{150mm}
\includegraphics[width=150mm,scale=1.5]{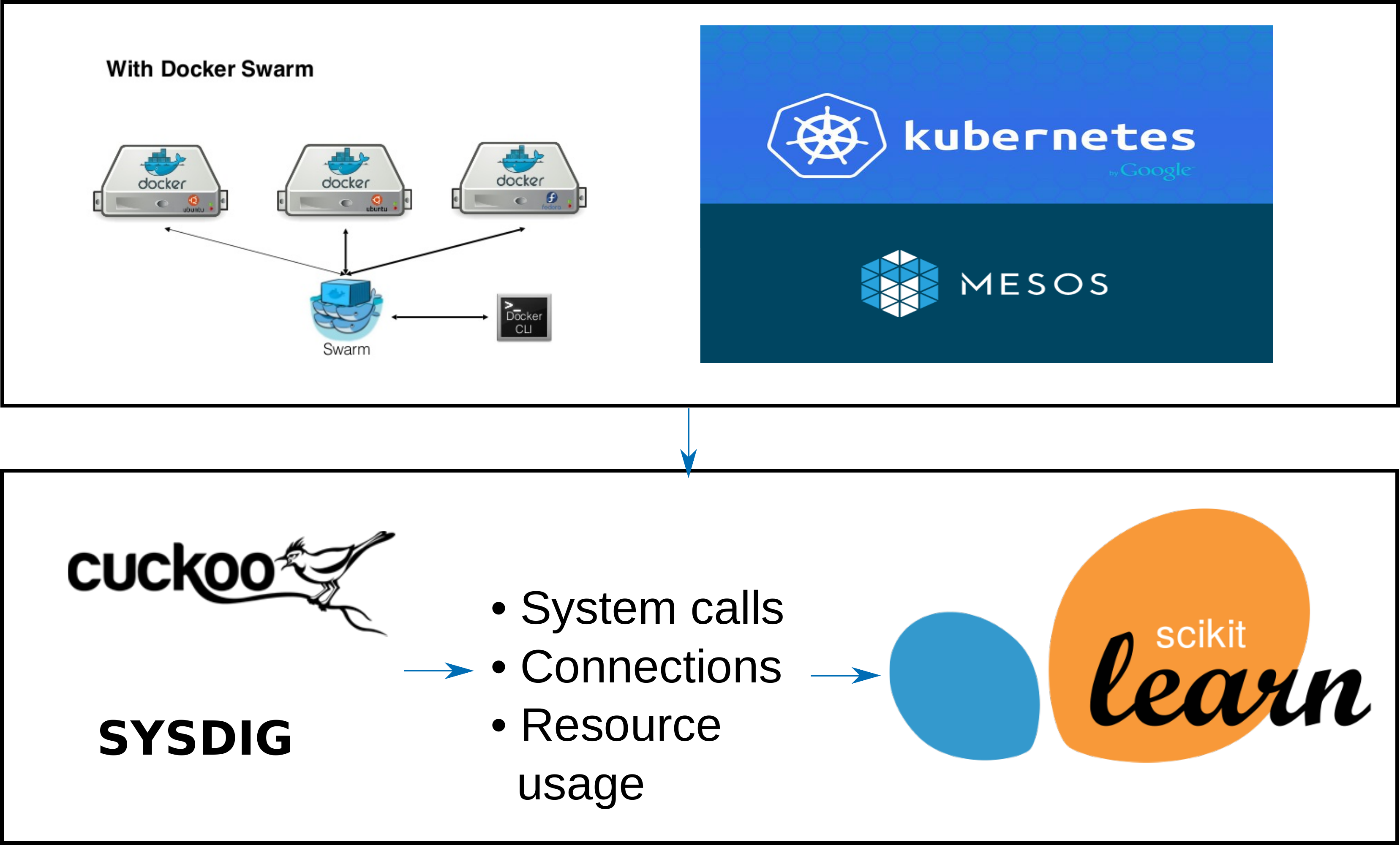}
\caption{\label{fig:figurepoc}Proof of concept implementation.}
\end{minipage}\hspace{2pc}%
\end{center}
\end{figure}

A network isolated machine was used for malware data collection. This
machine has the same setup as the Grid worker nodes. We have downloaded
a set of 10000 Linux malware samples from a security research web site
\cite{malware}. We ran the samples and collected the same information as for the
normal Jobs (logs, sequence of system calls, resource usage data). Finally we 
obtained a combined dataset that allows to train and test 
our selected Machine Learning algorithms. A representation of the implemented 
components are shown in the Figure~\ref{fig:figurepoc}.

\section{Conclusions}
\label{conclusions}
Distributed computing is a fundamental component of High Energy Physics
collaborations. Improving security in this kind of infrastructures requires
innovative tools to automatically detect security related incidents.
Security by isolation is also necessary to protect sensitive components allowing
traceability on Job activities. We propose the usage of Linux Containers in order 
to provide isolation without highly decreasing the expected performance.
We use Machine Learning techniques to provide generalization, overcoming common
IDS difficulties on finding even slightly different threats. We describe the
ongoing development process of a new security monitoring framework for Linux
Containers based HEP infrastructures. This is being tested as a proof of
concept for the ALICE experiment at CERN. We have collected a dataset of 
normal and malicious monitoring information from Grid Jobs and malware samples, 
that will be utilized to train and test ML algorithms. These
algorithms should enable autonomous Intrusion Detection and Prevention as an
important component of the proposed new framework. As future work we plan to
explore how our approach can be used to detect anomalies that go beyond the
security scope, for instance to find hardware failures or even human mistakes.

\ack
Authors acknowledge assistance from CERN security department specially Stefan Lueders and Romain Wartel.
This work is supported by the German Federal Ministry of Education and Research.

\end{document}